\documentclass[prb,twocolumn,showpacs,showkeys,preprintnumbers,floatfix,amsmath,amssymb,superscriptaddress]{revtex4}

\usepackage{graphicx}
\usepackage{bm}
\usepackage{epsfig}

\begin{document}

\title{Small-polaron coherent conduction in lightly doped ReTiO$_{3+
\delta/2}$ (Re=La or Nd)\\ thin films prepared by Pulsed Laser
Deposition}

\author{J. Li}
    \email{lijie@ssc.iphy.ac.cn}
    \affiliation{National Laboratory for Superconductivity,
Beijing National Laboratory for Condensed Matter Physics and
Institute of Physics, Chinese Academy of Sciences, Beijing 100080,
China.}

\author{F. B. Wang}
    \affiliation{National Laboratory for Superconductivity,
Beijing National Laboratory for Condensed Matter Physics and
Institute of Physics, Chinese Academy of Sciences, Beijing 100080,
China.}
    \affiliation{Sichuan University, College of Materials Science \& Engineering,
    Chengdu 610064, China}

\author{P. Wang}\author{M. J. Zhang}\author{H. Y. Tian}\author{D. N. Zheng}
    \affiliation{National Laboratory for Superconductivity,
Beijing National Laboratory for Condensed Matter Physics and
Institute of Physics, Chinese Academy of Sciences, Beijing 100080,
China.}

\date{\today}

\begin{abstract}

LaTiO$_{3+\delta/2}$, NdTiO$_{3+\delta/2}$, and
Nd$_{1-x}$Sr$_{x}$TiO$_{3+\delta/2}$ thin films have been
epitaxially grown on (100)SrTiO$_{3}$ and (100)LaAlO$_{3}$ single
crystal substrates by using the pulsed laser deposition technique.
The oxygen pressure during deposition has been carefully controlled
to ensure that the films are lightly doped in the metal side near
the metal-Mott-insulator boundary. A steep drop of the effective
carrier number at low temperatures has been observed in some of the
films, which may correspond to a gradually opening of the Spin
Density Wave (SDW) gap due to the antiferromagnetic spin
fluctuations. At elevated temperatures, a thermally induced decrease
of the Hall coefficient can also be clearly observed. In spite of
the fact that these films were prepared from different materials in
varied deposition conditions, the temperature dependence of their
resistance can all be almost perfectly fitted by a small-polaron
coherent conduction model
($R_s(T)=R_s(0)+C\omega_\alpha/\sinh^2(T_\omega/T)$). Careful
investigation on the fitting parameters implies that the frequency
of the phonon coupled to the electrons may be partially related to
the lattice distortion induced by the mismatch strain of the
substrates.
\end{abstract}


\maketitle

\section{Introduction}

The titanates ReTiO$_{3}$ (where Re is a trivalent rare-earth ions)
are canonical Mott-Hubbard type insulators with Ti$^{3+}$ (3$d^{1}$
$t_{2g}$) electron configuration. \cite{Imada} It has been widely
accepted that the insulating behavior is ascribed to the strong
Coulomb repulsion among the integer number of electrons at each Ti
sites, which opens a charge gap at the Fermi surface. The gap width
is determined by the electron correlation strength, $U/W$, where $U$
is the on-site Coulomb energy and $W$ is the one-electron bandwidth.
Varying the GdFeO$_{3}$-type distortion by changing the radius of Re
ions or by applying an external hydrostatic pressure, one can tune
the bandwidth $W$ and hence the correlation strength. A
bandwidth-controlled insulator-metal transition then can occur.
Alternatively, anomalous metallic states can also be derived from
the Mott insulators upon carrier doping, by substituting the
trivalent rare-earth with a divalent ion or by introducing excess
oxygen atoms into the lattice, the so-called filling-controlled
insulator-metal transition.

Generally speaking, the pure perovskite ReTiO$_{3}$ (Ti$^{3+},
3d^{1}$) crystals with a stoichiometric oxygen content can only be
acquired in an extremely reducing atmosphere.
\cite{Fujishima,Tokura} The Ti oxides have a strong tendency to
incorporate La and Ti vacancies, which are frequently denoted by an
excess of oxygen in the formula, ReTiO$_{3+\delta/2}$.
\cite{MacEachern} With the increase of nominal hole doping $\delta$,
the material evolves from a Mott insulator with a spin and orbital
ordering, to a paramagnetic metal of filling level $1-\delta$ in the
$3d$ band. As $\delta$ is larger than 0.4, the effective carrier
density in the system decreases, and additional oxygen layers appear
along the perovskite \{110\} planes. The end member ReTiO$_{3.5}$ is
a band insulator (Ti$^{4+}, 3d^{0}$) with a layer-perovskite
structure, which becomes ferroelectric at an enormously high
transition temperature. \cite{Lichtenberg}

The electronic properties of the Ti oxides in the vicinity of the
metal-Mott-insulator transition have been widely studied during the
past decade. A striking quadratic temperature dependence ($T^2$) of
resistivity has been conspicuously observed over a large temperature
range in the metallic LaTiO$_{3+\delta}$ \cite{Tokura} and
Sr-substituted Sr$_{1-x}$La$_{x}$TiO$_{3+\delta}$ \cite{Kumagai}
single crystals, accompanied by a remarkable electron effective mass
enhancement. This had been attributed to the strong
electron-electron correlation in the system. Other characteristic
properties, such as temperature independent Pauli-like
susceptibility and the $T$-linear electronic specific heat are also
discovered, the doped titanium oxides were then argued to be well
described by the Fermi liquid picture.

Nevertheless, in contrast to the comprehensive investigations on
single crystal samples, reports on titanate thin films
\cite{Ohtomo,Seo,Schmehl,Gariglio} are limited in quantity and the
analyses are somewhat at a superficial level. Hwang \textit{et al.}
\cite{Ohtomo} has demonstrated that, up to 6 unit-cell-thick
LaTiO$_{3}$ layers can be stabilized on SrTiO$_{3}$ substrate at
conventional deposition conditions. As the film grows thicker,
however, a high density of $\{110\}$ faults (excess oxygen layers
inserted) will develop, driven energetically to achieve the
equilibrium oxidative state of Ti. Accordingly most of the films
prepared in reduced oxygen pressures behave metallically, although a
diversity in the absolute resistivity value and the temperature
dependency was noticed. Among the few attempts to model the metallic
transport properties in titanate thin films, Gariglio \textit{et
al.} \cite{Gariglio} was very successful by considering the
interaction between electrons and phonons.

In this experiment, we deposited epitaxial thin films of different
titanates on both (100)SrTiO$_{3}$ (STO) and (100)LaAlO$_{3}$ (LAO)
single crystal substrates at varied temperatures and low oxygen
pressures. We are surprised to find that, the temperature
dependencies of the film resistivity can all be excellently fitted
by a small-polaron coherent conduction model \cite{Gariglio,Zhao}.

\section{Experiment}

The LaTiO$_{3+\delta/2}$ (LTO) and NdTiO$_{3+\delta/2}$ (NTO) thin
films have been grown on STO and LAO substrates by using the pulsed
laser deposition (PLD) technique in a high vacuum chamber. The
ceramic LaTiO$_{3.5}$ and NdTiO$_{3.5}$ targets were fabricated by
conventional solid-state reaction at 1370\r{}C and 1400\r{}C,
respectively, in air or flowing oxygen for 10 hours. The Sr doped
NTO film was prepared by sticking a small piece of STO single
crystal onto the ceramic NTO target. A KrF excimer laser with 248 nm
wavelength was employed for ablation. The substrate temperature was
within 650 to 850\r{}C, and the laser energy density was around 2
J/cm$^{2}$. Most of the films were deposited in the background
vacuum (BV) below 5$\times $10$^{ - 4}$ Pa, whereas a few were
prepared in slightly higher oxygen pressures. The deposition lasted
for 30 minutes with a laser repetition rate of 2 Hz. After growth
the films were annealed \textit{in situ} for 10 minutes at the
deposition temperature and oxygen pressure, and then cooled down to
room temperature in a ramp rate of 40\r{}C/min. The thickness of the
films is estimated to be $\sim$2000 {\AA}.

The crystal structure of these films was monitored \textit{in situ}
by a reflection high-energy electron diffraction (RHEED) system, and
studied \textit{ex situ} by x-ray diffraction. The electrical
resistivity and Hall coefficient were measured using the van der
Pauw geometry ($5\times5\rm mm^{2}$) in the temperature range 5
K$<\rm T<$300 K. The electrical contacts were made by indium
soldering. Magnetization measurements were carried out by first
cooling the samples down to 5 K in a 5 T magnetic field, and then
measuring in a warming run in 100 Oe. The Hall measurements were
carried out in an MPMS superconducting quantum interference (SQUID)
measuring system and a Maglab multipurpose magnetic measuring
system.

\section{Results and discussions}

\begin{figure}
\includegraphics{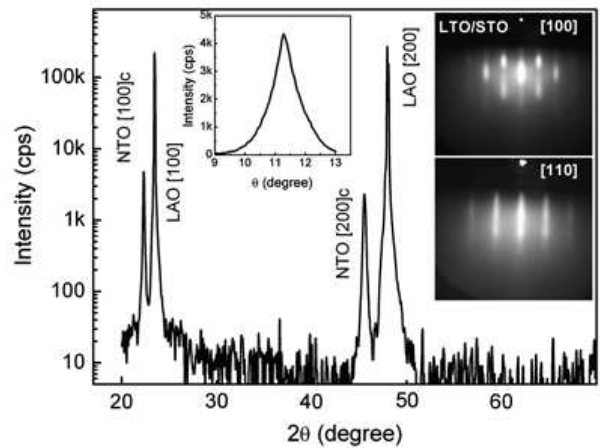}
\caption{\label{fig1}XRD \textit{$\theta$-2$\theta$} spectrum for
film NTO/LAO grown at 650\r{}C in vacuum. The inset is the rocking
curve of NTO[001]$\rm _{C}$. The $<$100$>$ and $<$110$>$ RHEED
images for film LTO/STO grown at 850\r{}C in vacuum is also
inserted.}
\end{figure}

All the films prepared are epitaxial, with excellent in-plane and
out-of-plane orientations. Shown in Fig. \ref{fig1} is a typical XRD
\textit{$\theta$-2$\theta$} spectrum for an NTO thin film deposited
on an LAO substrate at 650\r{}C in vacuum. No impurity phase can be
identified. The rocking curve of NTO[001]$\rm _{c}$ is inserted,
whose FWHM is $\sim $1\r{}. The XRD reflections for the films grown
on STO cannot be resolved from the substrate peaks due to the very
close lattice constants. Thereby, the RHEED images for an LTO thin
film grown on an STO substrate at 850\r{}C in vaccum is shown
instead, taken along the $<$100$>$ and $<$110$>$ zone-axes. The film
was grown in a cube-on-cube way, with its pseudo-cubic perovskite
{\{}001{\}} planes parallel to the film surface. The diffraction
streaks are bright and sharp, suggesting a good crystallinity of the
film. RHEED images for films on LAO substrates show almost no
dissimilarity. Some of the films, especially the NTO films, exhibit
a superstructure of $2a_{0}$ ($a_{0}$ is the lattice constant of the
cubic perovskite structure) in their RHEED diffraction patterns.
This is most probably due to the GdFeO$_{3}$-type distortion of the
perovskite lattice.

\begin{figure}
\includegraphics[width=.8\columnwidth]{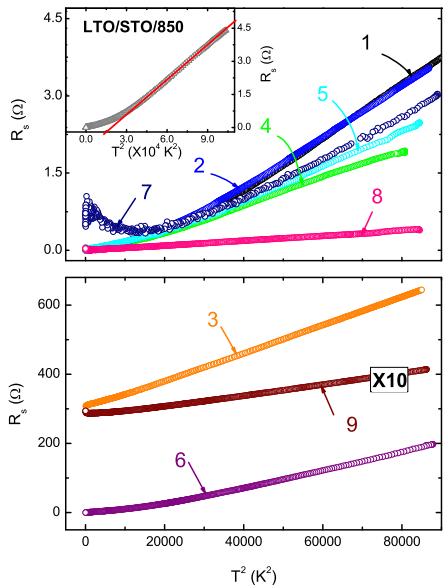}
\caption{\label{fig2}(Color online) $R_s$ versus $T^2$ curves for
the various lightly-doped epitaxial thin films deposited. The sample
details are listed in Table \protect\ref{table1}.}
\end{figure}

 The films deposited in the background vacuum or slightly higher oxygen pressures,
 say 10$^{-3}$ Pa, all behave like a metal from room temperature to 5 K. Some
shows a small resistivity upturn at low temperatures, while the
others do not unto 5 K. In Fig. \ref{fig2} we plot the
\textit{sheet} resistance $R_s$ versus $T^{2}$ curves for these
films. It is noticed that, although some of these samples have
resistance one or two orders of magnitude higher than the others,
these curves are similar in two ways: (a) in high temperature
regions, the resistance shows an approximately linear dependence on
$T^{2}$. (b) at a low temperature $T^*$, however, the resistance
starts to deviate visibly from this law and goes up, as can be
clearly seen in the inset, no matter if there is an upturn or not in
the $R-T$ curves. $T^*$ ranges from 120 K to 200 K or even higher.
The $T^{2}$ dependence of resistivity ($\rho=\rho_{0}+AT^{2}$) has
long been observed in transition metals \cite{Isshiki}, and later in
doped V$_{2}$O$_{3}$\cite{McWhan} and titanate single crystals
\cite{Tokura}. It has been attributed to the strong
electron-electron correlations in these systems. The coefficient $A$
in transition metal Iron, Nickel and Cobalt is in the order of
10$^{-11}\Omega$ cmK$^{-2}$, and that in the doped V$_{2}$O$_{3}$ is
5$\times$10$^{-8}\Omega$ cmK$^{-2}$. The $A$ values in our
experiment are determined to be between 9$\times$10$^{-11}\Omega$
cmK$^{-2}$ and 8$\times$10$^{-8}\Omega$ cmK$^{-2}$, comparable to
the previous results. Nevertheless, in the strong correlation frame,
the $T^{2}$ dependence is usually dominant at low temperatures. A
phonon scattering term shows up only as temperature increases.
Apparently, our data disagree with this rule, suggesting that the
conduction mechanism of epitaxial thin films may somewhat differ
from that of the single crystals.

\begin{figure}
\includegraphics[width=0.9\columnwidth]{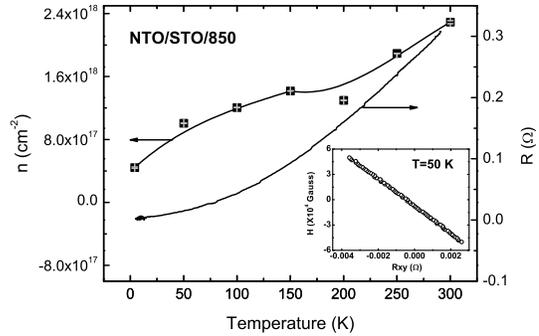}
\caption{\label{fig3}The $R_s-T$ and $n_s-T$ curves for an NTO thin
film grown on an STO substrate at 850\r{}C in vacuum.}
\end{figure}

On the other hand, while most of the previous studies reported
temperature-independent carrier density, which is a character of
Fermi liquid, our measurements demonstrate a clear temperature
variation of the Hall coefficient. In Fig. \ref{fig3}, we show the
$R_s-T$ curve for the NTO thin film grown on an STO substrate at
850\r{}C in a background vacuum. The sample represents the lowest
resistivity we have ever reached. Assuming that the film thickness
is 2000{\AA}, its room temperature resistivity is approximately
6.1$\mu\Omega$cm, and the value is only 4.2$\times
$10$^{-2}\mu\Omega$cm at 5 K. The Hall signal was measured in a
perpendicular magnetic field scanned from -5T to +5T at fixed
temperatures by using a Maglab-12 (Oxford). The field dependence of
the crossover resistance $R_{xy}$ was recorded. A typical $R_{xy}-H$
relationship is given in the inset of the figure, suggesting a
negative carrier type. The \textit{sheet} carrier density was read
from the slope of these curves using the standard free carrier Hall
effect formula $n_s=B/eR_{xy}$, and is also shown in the figure. The
error bar was added according to the linear fitting errors, which
are underestimated because the temperature fluctuations and the
geometric irregular of the sample are not considered. It is clear
that, although the sample is a good metal throughout the whole
temperature range measured, its carrier density decreases with
decreasing temperature. The carrier density $n$ is estimated to be
from 2.2$\times $10$^{22}$cm$^{-3}$ to 1.1$\times
$10$^{23}$cm$^{-3}$, which is, however, unphysical since the highest
$n$ calculated by assuming one $d$ electron per unit cell is only
around 1.65$\times $10$^{22}$cm$^{-3}$. The overestimated carrier
density may be originated from experimental errors, or most probably
from the simple free carrier assumption.

\begin{figure}
\includegraphics[width=0.9\columnwidth]{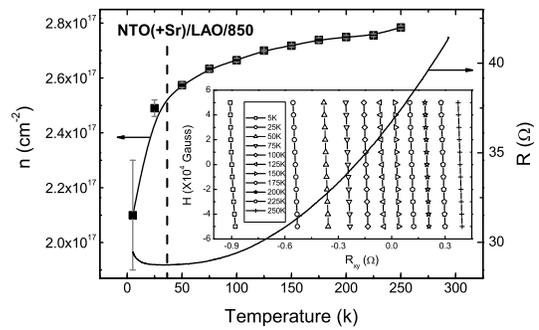}
\caption{\label{fig4}The $R_s-T$ and $n_s-T$ curves for an Sr doped
NTO thin film grown on an LAO substrate at 850\r{}C in vacuum. The
dash line denotes the M-I transition temperature.}
\end{figure}

\begin{table*}
\caption{\label{table1}}
\begin{ruledtabular}
\begin{tabular}{cccccccccc}
 \\
 No. & Sample & Denotation & P$\rm _{O}/BV$ & $R_s$(0) & $C\omega_\alpha$ & $C$ & $T_\omega$ & $\chi^{2}$ & $R^{2}$ \\
  & & & (Pa) &($\Omega$) & ($\Omega$) & ($10^{-14}\Omega\cdot \rm S^{-1}$) & (K) \\\hline
 \\
 1 & 260505 & LTO/STO/850 & 5e-4 & 0.024(1) & 4.47(4) & 6.0 & 284.6(9) & 0.00086 & 0.99952\\
 2 & 120905 & LTO/STO/650 & 4.3e-4 & 0.041(1) & 3.49(3) & 5.1 & 260.0(9) & 0.00055 & 0.99959\\
 3 & 230505 & LTO/LAO/850\footnotemark[1] & 5e-4 & 312.0(1) & 38.8(9) & 151.7 & 97.90(8) & 1.67078 & 0.99989\\
 4 & 111104 & LTO/STO/780 & 4.5e-4 & 0.021(2) & 0.85(2) & 1.8 & 176.65(5) & 0.00042 & 0.99905\\
 5 & 191104 & LTO/STO/650 & 5e-4 & 0.053(2) & 1.78(4) & 3.0 & 226.62(5) & 0.00067 & 0.99907\\
 6 & 140805 & NTO/STO/850 & 6e-3 & 1.80(9) & 143.1(7) & 237.0 & 230.77(9) & 1.89505 & 0.99923\\
 7 & 121005 & NTO/STO/650 & 5e-4 & 0.320(5) & 7.0(2) & 7.1 & 374.6(4) & 0.00073 & 0.99897\\
 8 & 120805 & NTO/STO/850 \footnotemark[2]& 6e-4 & 0.0016(1) & 0.0005(4) & 0.008 & 25.0(9) & 1.0273e-6 & 0.998\\
 9 & 241005 & NTO(+Sr)/LAO/850& 2.4e-4 & 28.798(3) & 5.77(3) & 11.8 & 186.3(5)& 0.0011 & 0.99993\\
\end{tabular}
\end{ruledtabular}
\footnotetext[1]{This film can also be well fitted by the $T^{2}$
relationship.} \footnotetext[2]{Fitting of this sample gives a low
reliability.}
\end{table*}

Shown in Fig. \ref{fig4} is the Sr doped NTO on an LAO substrate at
850\r{}C in a background vacuum. This sample shows a resistance two
orders of magnitude higher and a clear M-I transition at $\sim$20 K.
The Hall measurement for this sample was performed by a SQUID
MPMS£­5 (Oxford), and also in a field of -5T to +5T. Because of the
poor alignment of the voltage electrodes, the $R_xy-H$ curves
measured are severely asymmetric for positive and negative fields,
but the slopes of these curves are distinct, as can be seen in the
inset. The derived carrier density decreases gradually with
decreasing temperature, and drops steeply near the M-I transition
point, resulting in the upturn of resistivity at low temperatures.
Here we note that, for Sr doped NTO single crystals, the low
temperature antiferromagnetism (AFM) survives until the doping level
is higher than 0.2 \cite{Katsufuji}. Strong antiferromagnetic
fluctuations may persist well above T$\rm _{N}$, which gradually
opens a spin density wave (SDW) gap over some portion of the Fermi
surface as the temperature is decreased \cite{Taguchi}. We believe
this is the reason for the effective carrier number collapse and the
concomitant resistivity increase below 20 K. Inductively coupled
plasma - atomic emission spectrometry (ICP-AES) reveals that,
however, the La:Sr ratio in this sample is 1:0.93, which means that
the filling level $1-x-\delta$ should be around 0.5. Because of the
substrate effect, as we will discuss later, AFM may still lingering
in the sample, although the chemical analysis destroys the sample,
making the magnetization measurement unavailable. The carrier
density $n$ is derived to be from 1.0$\times $10$^{22}$cm$^{-3}$ to
1.4$\times $10$^{22}$cm$^{-3}$, while the ideal carrier density
should be no more than 8.5$\times $10$^{21}$cm$^{-3}$. Therefore,
the derived carrier density in this sample is again overestimated.

Therefore, although the two samples show different transport
behaviors, they all exhibit a thermally induced increase of the
effective carrier number, which is inconsistent with the Fermi
liquid picture. Considering our $R-T$ fitting results, as will be
detailed below, we prefer to explain this phenomenon using a polaron
conduction picture. A theory of small-polaron coherent conduction
was developed in 1968 by Bogomolov \textit{et al.} \cite{Bogomolov},
and has been adopted by Zhao \textit{et al.} \cite{Zhao} to explain
the metallic transport behavior in the doped perovskite manganite
thin films. Gariglio \textit{et al.} \cite{Gariglio} also found
recently that this model could interpret well the conduction
behavior of their LTO thin films grown on LAO substrates. In this
model, the resistivity is given by

\[
\rho(T)=\rho_0+(\hbar^2 /ne^2a^2t_p)\sum\limits_\alpha
{A_\alpha\omega_\alpha/\sinh ^2(\hbar\omega_\alpha/2k_BT)},
\]

\noindent where $n$ is the carrier density, $a$ is the lattice
constant, $t_p$ is the effective hopping integral of polarons,
$\omega _\alpha$ is the average frequency of one optical phonon
mode, and $A_\alpha$ is a constant depending on the electron-phonon
coupling strength. The fully activated heavy particles are scattered
by thermal phonons and thus the resistivity increases with
increasing temperature. Assuming that only the low-lying optical
mode with a strong electron-phonon coupling contributes to the
resistivity, we fit the $R-T$ curves shown in Fig. \ref{fig2} using
the expression

\[
R_s(T)= R_s(0)+ C\omega_\alpha/\sinh^2(T_\omega/T).
\]

\noindent The fitting parameters are listed in Table \ref{table1}.
The curve fittings were done without weighting. For the curves with
an upturn at low temperatures, we fit only the data above the
transition point. It is striking that the model can perfectly fit
almost all of the experimental data in Fig. \ref{fig2}, in spit of
the fact that these samples are made from different kinds of
materials on both STO and LAO substrates at varied temperatures and
in different oxygen pressures. Values of the \textit{goodness of
fit} $R^{2}$ are all $\ge$ 0.999, except that for Sample 120805
(also shown in Fig. \ref{fig3}). $R^{2}$ for this sample is slightly
lower, most probably due to its rather high single-to-noise ratio.
Therefore the fitting result is ignored in the following analyses.
The excellent agreement between the experimental and the calculated
curves strongly implies the presence of small-polarons and their
metallic conduction in the perovskite titanate thin films. The small
fitting error may have its origin that, in the model the carrier
density $n$ is treated as a constant, while in fact it changes
slowly with temperature.

\begin{figure}
\includegraphics[width=1.05\columnwidth]{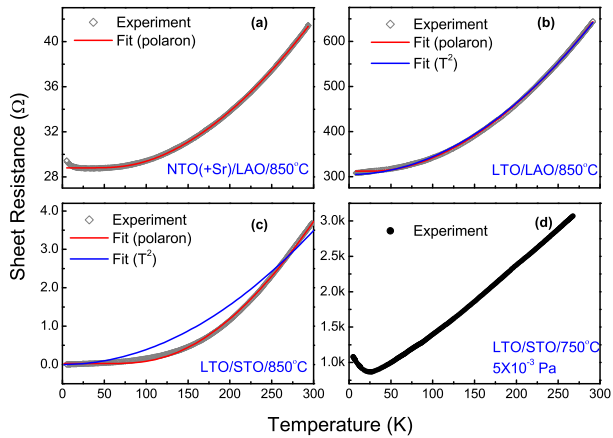}
\caption{\label{fig5}(Color online) Examples of the small-polaron
fitting results.}
\end{figure}

In Fig. \ref{fig5}(a), (b), and (c) we give three examples of the
curve fittings. Fig. \ref{fig5}(a) is for curve 9 (sample 241005,
also shown in Fig. \ref{fig4}). Fitting was performed only from 35 K
to 295 K, but the calculated curve is extended to low temperatures
in the figure. In Fig. \ref{fig5}(b) we show fitting for curve 3
(sample 230505) using both the $T^{2}$ relationship and the
small-polaron model. This sample shows the best fitting result by
the $T^{2}$ relationship among all the films prepared, even though
the $R^{2}$ value for small-polaron is still larger. Curves for the
other films presented in Fig. \ref{fig2} can only be satisfactorily
fitted by the polaron model, as illustrated in Fig. \ref{fig5}(c)
(curve 9, sample 260505). Nevertheless, only the films with hole
doping near the Mott-insulator-metal transition boundary can be well
fitted by the small-polaron coherent conduction model. For thin
films deposited in elevated oxygen pressures, as that shown in Fig.
\ref{fig5}(d), though still behaves like a metal, at high
temperatures the resistivity is remarkably enhanced and almost
temperature linearly dependent. This sample corresponds to a highly
oxygen-doped state far away from the I-M transition. With increasing
of the oxygen pressure, extra oxygen layers gradually appear in the
originally cubic perovskite lattice. The film then becomes a mixture
of two-dimensional (2D) and three-dimensional (3D) regions. The
linearly T-dependent behavior may suggest a kind of 2D conduction
behavior, like in the case of high-temperature superconductor normal
state \cite{Gariglio}.

\begin{figure}
\includegraphics[width=.88\columnwidth]{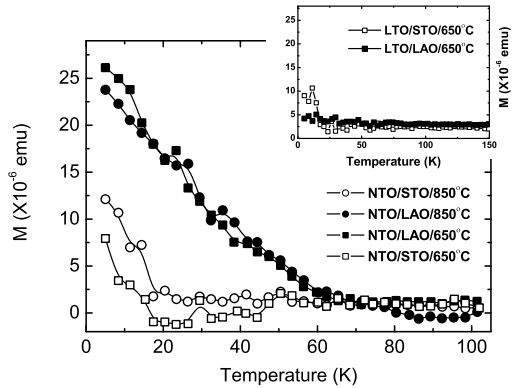}
\caption{\label{fig6} Magnetization curves for some of the LTO and
NTO thin films measured in a perpendicular magnetic field of 100 Oe
(5T field-cooled ) \protect\cite{Katsufuji}.}
\end{figure}

By scrutinizing Table 1 one can find: (a) While the LTO films grown
on STO substrates show very small residual resistance $R(0)$, the
value is slightly higher for NTO films on STO, and is the largest
when both films are deposited on LAO substrates. It is known that
the nominal hole concentration required to completely suppress the
low-temperature AFM ordering increases with the correlation strength
$U/W$ \cite{Katsufuji}. Since the bandwidth $W$ in NTO is narrower
than that in LTO and $U$ in the two systems are equivalent, with
similar doping, i.e. deposited in identical conditions, the AFM
ordering still lingers in the NTO thin films, but dies in the LTO
films, as clearly verified in Fig. \ref{fig6}. The figure also
indicates that the NTO thin films grown on LAO substrates have a
much higher $T_N$ and a larger magnetic moment than the films grown
on STO. This suggests that the former are `` much more correlated
''. Further research relevant to the substrate effect will be
reported elsewhere later. The AFM ordering and fluctuations open an
SDW gap on the Fermi surface, which decreases the Fermi surface
area. The residual resistivity $R_s(0)$ is in inverse proportion to
the Fermi surface area \cite{Taguchi}. Thus the residual resistance
lifts with increasing electron correlation strength. (b) The
constant $C\sim A_\alpha\hbar^2/ne^2a^2t_p$ manifests its largest
value in sample 140805, which was deposited in a higher oxygen
pressure and concomitantly with a lower filling $n\approx 1-\delta$.
The second high value is found in sample 230505, a film grown on an
LAO substrate. Intuitionally, $C$ is roughly proportional to
$m^*/n$, where $m^*$ is the effective mass of polarons, and $n$ is
the carrier density. Apparently, polarons in the films on LAO
possesses a larger $m^*$ due to the higher correlation strength. One
may note that, the heavily doped sample 241005 only exhibits an
intermediate $C$, where the effect of its low $n$ perhaps has been
compensated by a remarkably reduced electron correlation, since the
film is already away from the Mott-insulator-metal transition. (c)
The characteristic temperature $T_\omega$ of the phonon coupled to
electrons disperses in a scope from 97 to 374 K. This, however,
disagrees with Gariglio's \cite{Gariglio} argument that only a
dominant optical phonon mode of $\hbar\omega_{0}/k_{B}=80$ K
contributes substantially to the resistivity. On the contrary,
$T_{\omega}$ shows a weak dependence on the substrate materials. The
phonon frequency for the films grown on STO tends to be higher than
that on LAO. This may suggest a flaw in attributing the phonon
simply to the tilt/rotation of the oxygen octahedra in perovskite
materials. The lattice mismatch strain between the films and the
substrates may also play an important role, which directly affects
the film lattice distortion, and thus the phonon. Further studies
are required to clarify this issue.

\section{Conclusions}
In summary, we have deposited LTO, NTO, and NSTO thin films on
(100)STO and (100)LAO single crystal substrates by using the PLD
technique. The oxygen pressure during deposition has been carefully
controlled to ensure that the films are lightly hole doped near the
metal-Mott-insulator boundary. The films are all epitaxially grown
in a cube-on-cube way, and are metallic from room temperature to
liquid helium. Their resistivity is $T^2$ dependent at high
temperatures, but deviates from this law at temperatures below
$T^*$. A temperature dependent Hall coefficient has also been
observed. Therefore the films cannot fit in with the Fermi liquid
frame. In spite of the fact that these films were prepared from
different materials in varied deposition conditions, the temperature
dependence of their resistivity can all be nearly perfectly fitted
by a small-polaron coherent conduction model. In some of the films,
at low temperatures a steep drop of the effective carrier density
has been observed, which suggests an opening of the SDW gap due to
the AFM spin fluctuations. Concomitantly the Fermi surface area
decreases, so that these films have a relatively high residual
resistance $R_s(0)$. Careful investigation on the fitting parameters
implies that the substrates may have an important effect on the
electron correlation strength in the films, and also on the
frequency of the phonon coupled to the electrons.

\section{Acknowledgement}

The project is sponsored by the National Natural Science Foundation
of China under grant Nos. 50472076, 10574154, and 10221002, and the
Ministry of Science and Technology, China (2006CB601007). The
authors would like to thank Mr. H. Yang for the Hall effect
measurements.

\end{document}